\documentclass[prl,superscriptaddress,reprint,showpacs]{revtex4-1}
\usepackage{siunitx}
\usepackage{color}
\usepackage{graphicx,textcomp,amssymb,amsmath,dcolumn,hyperref}
\usepackage[style=british]{csquotes}

\begin{document}

\title{Shell Filling and Trigonal Warping in Graphene Quantum Dots}

\author{R.~Garreis}
\email{garreisr@phys.ethz.ch}

\affiliation{ETH Zurich (Swiss Federal Institute of Technology in Zurich), 8093 Zurich, Switzerland}

\author{A.~Knothe}
\affiliation{National Graphene Institute, University of Manchester, Manchester M13 9PL, United Kingdom}
\author{C.~Tong}
\author{M.~Eich}
\author{C.~Gold}


\affiliation{ETH Zurich (Swiss Federal Institute of Technology in Zurich), 8093 Zurich, Switzerland}

\author{K.~Watanabe}
\author{T.~Taniguchi}
\affiliation{National Institute for Material Science, 1-1 Namiki, Tsukuba 305-0044, Japan}
\author{V.~Fal'ko}
\affiliation{National Graphene Institute, University of Manchester, Manchester M13 9PL, United Kingdom}
\affiliation{Henry Royce Institute for Advanced Materials, M13 9PL, Manchester, United Kingdom}
\author{T.~Ihn}
\author{K.~Ensslin}
\author{A.~Kurzmann}
\affiliation{ETH Zurich (Swiss Federal Institute of Technology in Zurich), 8093 Zurich, Switzerland}

\date{\today}

\begin{abstract}
Transport measurements through a few-electron circular quantum dot in bilayer graphene display bunching of the conductance resonances in groups of four, eight and twelve. This is in accordance with the spin and valley degeneracies in bilayer graphene and an additional threefold 'minivalley degeneracy' caused by trigonal warping.
For small electron numbers, implying a small dot size and a small displacement field, a two-dimensional s- and then a p-shell are successively filled with four and eight electrons, respectively. For electron numbers larger than 12, as the dot size and the displacement field increase, the single-particle ground state evolves into a three-fold degenerate minivalley ground state. A transition between these regimes is observed in our measurements and can be described by band-structure calculations.
Measurements in magnetic field confirm Hund's second rule for spin filling of the quantum dot levels, emphasizing the importance of exchange interaction effects.
\end{abstract}
	
\maketitle
	
Few-electron quantum dots have been studied in various semiconductors, such as InGaAs \cite{Tarucha1996, Kouwenhoven2001}, GaAs \cite{Ciorga2000}, InAs \cite{Shorubalko2006,Fuhrer2007}, or silicon \cite{Wang1994,Culcer2009,Leon2020}. Investigation of their ground and excited states, and of their addition spectra led to a comprehensive understanding of orbital and spin degeneracies, and hence enabled the implementation of solid state qubits \cite{Loss1998,Zhang2018,Wang2018,Cogan2018}. For vertical quantum dots etched into a circular geometry, shell filling and spin filling according to Hund's rules was observed \cite{Tarucha1996}.

A relatively new and promising material for quantum dot qubits is graphene \cite{Trauzettel2007}. Almost 99\% of the carbon atoms have zero net nuclear spin reducing hyperfine interactions compared to III-V semiconductors. Furthermore, carbon is a light element with reduced spin-orbit effects even compared to silicon \cite{Hongki2006}. These properties promise coherence times for qubits in graphene exceeding those of current semiconductor qubits. Due to the recent improvements in fabrication techniques for graphene nanostructures, few-electron or -hole quantum dots have been realized in bilayer graphene
\cite{Overweg2018,Eich2018, Banszerus2018,Kurzmann2019Chargdet,Kurzmann2019,Banszerus2020,Eich2020,banszerus2020electronhole,tong2020} that are comparable in quality to the best devices in GaAs.

However, the quantum dots' ground and excited states, and their addition spectra are not yet fully understood. Charge carriers in large-area bilayer graphene devices possess two-fold valley and two-fold spin degrees of freedom, as well as a non-trivial minivalley bandstructure due to trigonal warping \cite{McCann2007bandstructure,Xiao2010,Varlet2015,Knothe2018}. Increasing the displacement field perpendicular to the bilayer graphene sheet increases the induced bandgap and enhances the depth of the three minivalleys formed around the K and K' points \cite{Varlet2015, Knothe2018, Overweg2018}.
Relevance of these minivalleys for low-energy states in quantum dots has so far been predicted only theoretically \cite{Knothe2020}.

Here, we experimentally investigate the effects of trigonal warping in a nearly circular quantum dot in bilayer graphene. Starting from the empty quantum dot we observe a successive bunching of four, eight and twelve conductance resonances. We attribute these bunchings to the transition from a level scheme given by two-dimensional s- and p-shells for the first electrons, to a level scheme dominated by mini-valleys with three-fold degeneracy for more than twelve electrons. Theoretical band structure calculations confirm this transition and are in good agreement with our experimental observations. The circularity, the size and the band gap of the quantum dot are calculated using self-consistent \textsc{Comsol Multiphysics} simulations for the potential landscape and a capacitive tight-binding model. Measurements in a magnetic field applied parallel to the graphene plane show, that spin-filling into nearly degenerate levels obey Hund's second rule.

\begin{figure}
	\includegraphics[width=00.48\textwidth]{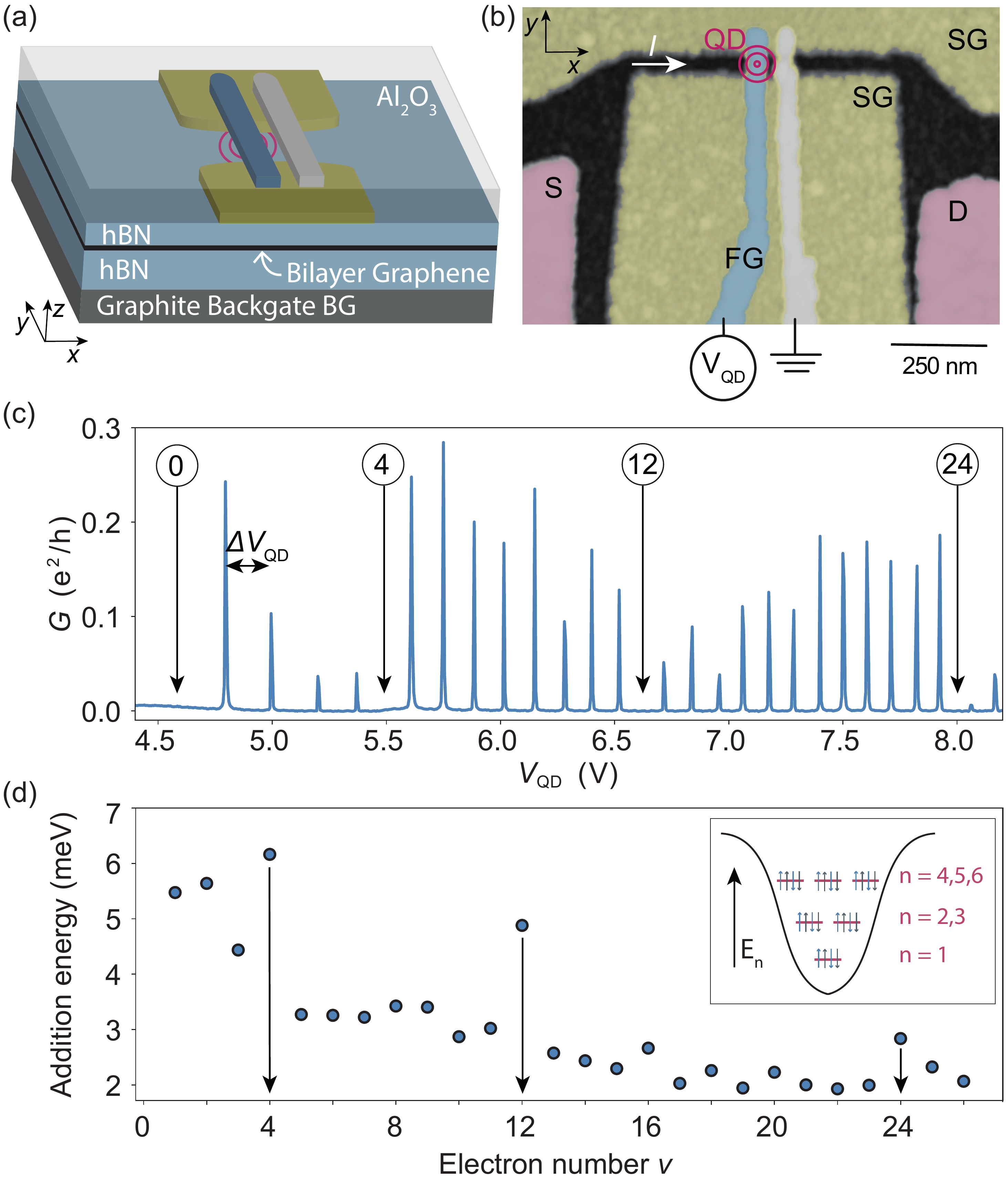}
	\caption{(a) Schematic representation of the stack. From bottom to top it is built up with a graphite back gate (grey), a bottom hBN (light blue), a bilayer graphene flake capped with a top hBN. Separated vertically by an aluminium oxide layer, the split gates (gold) and finger gate (dark blue) are used to electrostatically define a quantum dot (depicted in red). (b) False color atomic force microscope picture of the sample, with source (S) and drain (D) contacts (rose), split gates (SG, gold) and finger gate (FG, blue). The second finger gate (grey) is grounded. (c) Two terminal conductance trace through the dot with respect to the applied finger gate voltage $V_{\mathrm{QD}}$. (d) Addition energy needed for an extra electron versus number of electrons $\nu$ in the dot extracted from (b). Insert: Schematic depiction of the potential well including the energy levels $n$ corresponding to the measured bunching, each occupied with four electrons.}
	\label{fig1}
\end{figure}

The fabrication of the heterostructure shown schematically in Fig.~\ref{fig1}(a) follows the general procedure described in previous publications \cite{Wang2013, Overweg2018, Eich2018, Banszerus2018}. However, here, the thickness of the hBN layers (bottom: \SI{28}{nm}, top: \SI{34}{nm}), the split gate separation ($\SI{100}{nm}$), the thickness of the aluminium oxide layer ($\SI{30}{nm}$) as well as the width of the finger gate ($\SI{20}{nm}$) yield a rather circular shape of the confinement potential as demonstrated later by simulations [see inset Fig.~\ref{fig_theo_exp}(a)]. Figure~\ref{fig1}(b) shows a false color atomic force microscope image of the two layers of metal gates fabricated on top of the heterostructure. The split gates [golden in Fig.~\ref{fig1}(a),(b)] are used to form a conducting channel (black in Fig.~\ref{fig1}b)) \cite{Overweg2018}. For the measurements discussed in this paper the right finger gate [gray in Fig.~\ref{fig1}(b)] is grounded while the left [blue in Fig.~\ref{fig1}(b)] is used to form a quantum dot (QD) underneath it \cite{Eich2018,Banszerus2018}. Unless stated otherwise, a p-type conducting channel is formed between the two split gates, with insulating regions below them, by applying $V_{\mathrm{BG}} = \SI{-2.3}{V}$ to the back gate, and $V_{\mathrm{SG}} = \SI{1.242}{V}$ to the split gate. The finger gate is then used to form an n-type quantum dot, where the p-n junctions forming between the dot and the channel act as tunnel barriers \cite{Eich2018}. Conductance measurements are taken at an electron temperature of $\SI{\sim100}{mK}$ by applying a symmetric DC bias of \SI{100}{\micro V} and measuring the current in a two-terminal setup. 

The single-particle spectrum of this quantum dot is investigated using addition spectroscopy \cite{Tarucha1996, Leon2020}.
Figure~\ref{fig1}(c) shows the conductance through the quantum dot as a function of finger gate voltage $V_\mathrm{QD}$. At $V_{\mathrm{QD}} = \SI{4.8}{V}$, the first electron is loaded into the dot, as confirmed with a charge detector neighboring the dot (shown in \cite{Kurzmann2019Chargdet}), and more electrons follow for higher voltages. Bunching of successive Coulomb resonances into groups of four, eight and twelve is observed. The separation between neighboring Coulomb resonances reflects the energy needed to load the next electron into the dot. In a model based on the Hartree-approximation, it is the sum of the charging energy and the separation between the lowest unoccupied and the highest occupied single-particle level  \cite{Tarucha1996,Eich2018}.

With finite bias measurements (presented in Fig.\,S1 in the supplemental material \cite{SM_all}) the finger-gate lever arm $\alpha = 0.027$ is determined for the first electron. It decreases to $\alpha = 0.019$ for the twenty-forth electron due to increasing dot size. These lever arms allow us to convert the gate voltage differences $\Delta V_{\text{QD}}$ [horizontal axis of Fig.~\ref{fig1}(c)] into energy differences $\Delta E=e\alpha\Delta V_\mathrm{QD}$ \cite{Tarucha1996}. For the subsequent analysis, the lever arm is determined individually for each electron number.

We plot the addition energy for successive filling of electrons in Fig.~\ref{fig1}(d). For the first electron the addition energy is $\SI{5.5}{meV}$. It generally decreases with increasing number of electrons in the dot \cite{Tarucha1996,Eich2018}, as the electronic size of the dot increases. When the quantum dot is filled with the fourth, the twelfth and the twenty-fourth electron, an enhanced addition energy is observed, which indicates shell filling in the quantum dot as we will further confirm below \cite{Tarucha1996}. 

\begin{figure}
	\includegraphics[width=0.48 \textwidth]{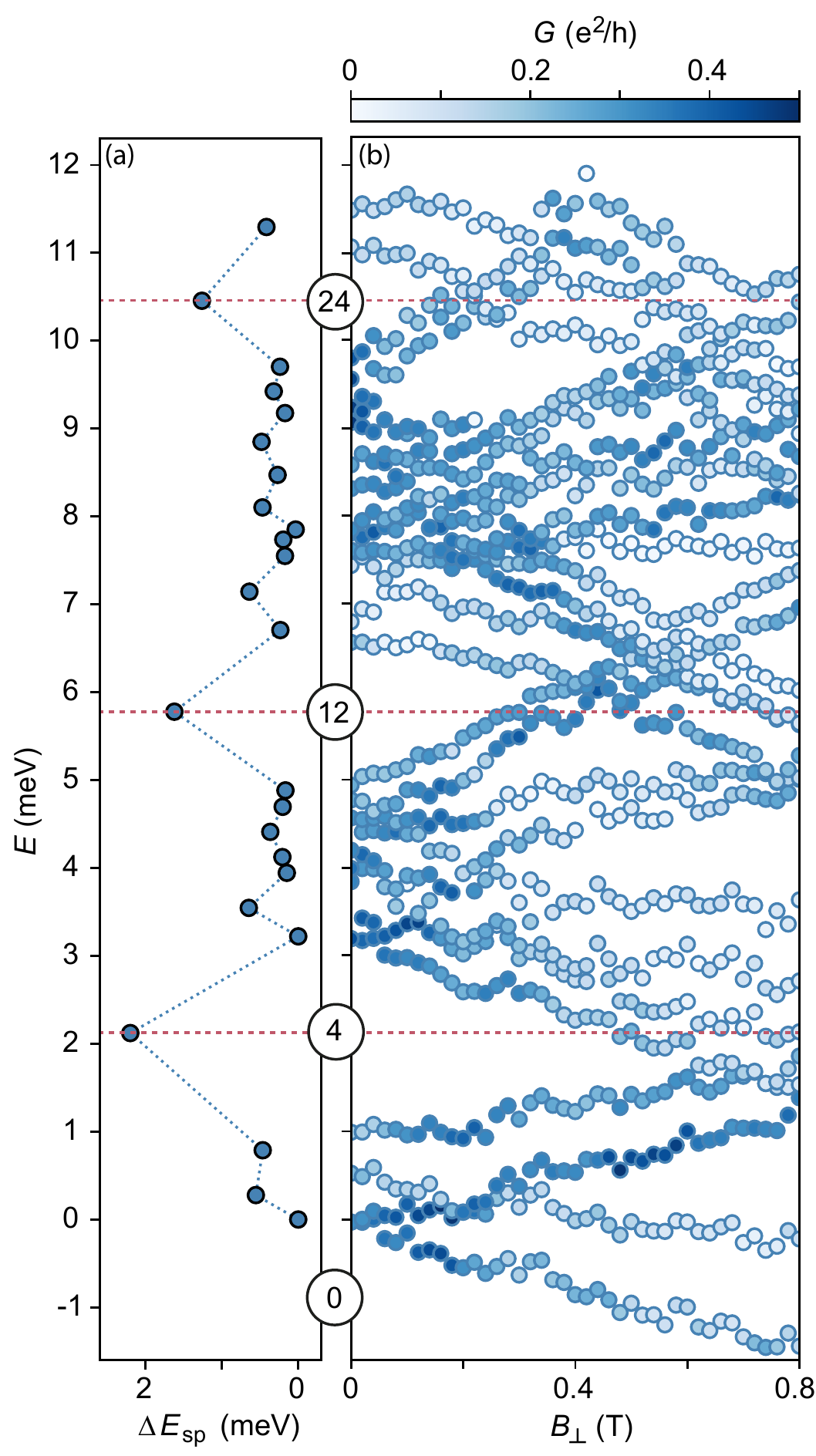}
	\caption{(a) Single particle level spacing of the quantum dot as a function of the energy scale extracted from the single particle energy dispersion in (b) at zero field. Corresponding number of electrons in the dot is marked. (b) Single particle energy dispersion with perpendicular magnetic field $B_{\perp}$, taken at $V_{\mathrm{BG}} = \SI{-2.5}{V}$ and $V_{\mathrm{SG}} = \SI{1.397}{V}$.}
	\label{fig_Bperp}
\end{figure}

\begin{figure*}
	\includegraphics[width=0.9 \textwidth]{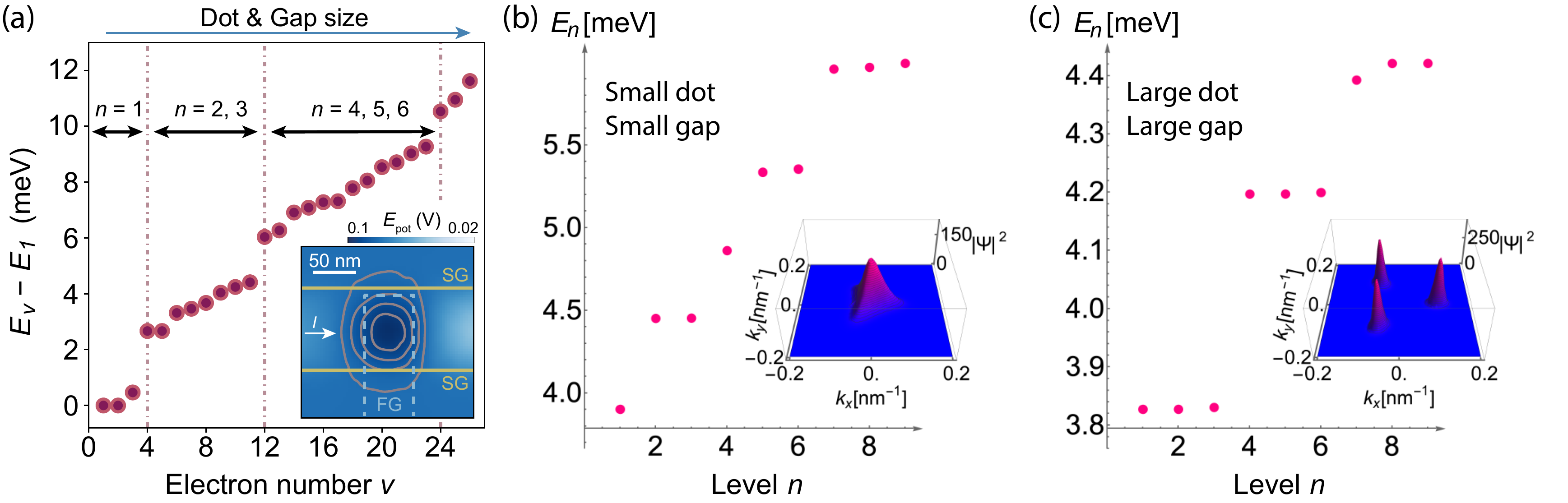}
	\caption{(a) Single particle energy levels extracted from Fig.~\ref{fig_Bperp}. The measured spectrum shows that the second and third (n= 2,3) energy levels are nearly degenerate. The same holds true for the fourth, fifth, and sixth levels (n= 4,5,6). Inset: Potential landscape induced to the bilayer graphene to form a quantum dot, where  $V_{\mathrm{QD}} = \SI{6.3}{V}$, calculated with \textsc{Comsol Multiphysics}. (b)-(c) Calculated quantum dot spectra for a theoretical model of a rotationally symmetric quantum dot (each level four-fold degenerate in the spin and valley degree of freedom). With increasing dot and gap size, multiplicity of the levels changes and orbital triplet degeneracies emerge due to the three minivalleys around each of bilayer graphene's valleys. Values for small (large) dot and gap sizes for the calculations were chosen to be similar to the experimentally determined values. Furthermore, Fig.\,S3 in the supplemental material shows the robustness of this transition over a wide range of parameters \cite{SM_all}. The insets show the momentum space probability distribution of the lowest dot state in the $K^+$ valley.}
	\label{fig_theo_exp}
\end{figure*}

We obtain information about the magnetic properties of the individual single-particle energy levels by measuring their response to an external magnetic field applied perpendicular to the graphene sheet. Subtracting the charging energy between neighboring resonances yields the magnetic field dispersion of the single particle energy levels $E_\nu$ shown in Fig.~\ref{fig_Bperp}(b) \cite{Duncan2000}. The two prominent slopes with opposite sign of the levels as a function of $B_\perp$ seen in Fig.~\ref{fig_Bperp}(b) are well-known \cite{Eich2018} and reflect the valley splitting due to the opposite magnetization of the K- and K'-valley states. An approximate zero-field single-particle level spacing $\Delta E_{\text{sp}}$ [Fig.~\ref{fig_Bperp}(a)] is extracted using the charging energy for each fixed electron number, assuming reasonably that charging energy is independent of magnetic field \cite{Duncan2000}. We observe again increased level spacings at four, twelve, and twenty-four electrons, same as the addition energy in Fig.~\ref{fig1}(d), which confirms the shell filling interpretation.

To estimate the circularity of the quantum dot we evaluate its size and shape by solving Poisson's equation in three dimensions with an electron density in the graphene plane determined self-consistently within the Thomas-Fermi approximation. Details of these evaluations are shown in the Supplemental Material \cite{SM_all}. Inset of Fig.~\ref{fig_theo_exp}(a) shows the simulated potential landscape for an exemplary finger gate voltage. The yellow lines outline the split gates forming the conducting channel, where the light blue dashed lines delineate the finger gate that is used to form the dot. Below the finger gate, the higher electric potential forms the potential well serving as the quantum dot confinement potential. The grey lines are equipotential lines indicating the shape of the dot confinement. The nearly circular shape of the equi-potentials may cause approximate orbital level degeneracies.

For a comparison of the experimentally determined single-particle level spacing in Fig.~\ref{fig_Bperp}(a) with the theoretical calculations modeling the system for spin-less electrons in a single valley, we
plot $E_{\nu}-E_1$, the single-particle energy levels $E_\nu$ with respect to the first energy level $E_1$, in Fig.~\ref{fig_theo_exp}(a). The measured spectrum shows that the single particle energies of the first four, fifth to twelfth and the thirteenth to twenty-fourth electron are each closer together than the energy separation to the next bunch of single particle energies. This implies that the second and third energy level ($n=2,3$) of a given spin and valley are nearly degenerate. The same holds for the fourth, fifth, and sixth levels ($n=4,5,6$) [see insert of Fig.~\ref{fig1}(d)].

The three-fold degeneracy obtained in the Darwin-Fock model for the third orbital level ($n=4,5,6$) is accidental because of the parabolic confinement potential and cannot explain our observations. In addition, the Berry curvature and the resulting orbital magnetic moment in bilayer graphene lift the degeneracy between d-shell states with different radial quantum numbers \cite{Knothe2020}.

Figures~\ref{fig_theo_exp}(b) and (c) show calculated energy levels in a circular bilayer graphene quantum dot for different band gaps inside the dot and different dot sizes, where each data point represents four degenerate spin and valley states.
We theoretically describe the quantum dot by a smooth, rotationally symmetric confinement potential with a spatially varying spectral gap (see Supplemental Material for details about the model and the calculation \cite{SM_all}). We model the experiment by simultaneously changing the dot size, $L$, and the gap inside the dot (as they are tuned by the finger gate) while keeping the gap under the split gates constant.
In Fig.~\ref{fig_theo_exp}(b) we see that for a small dot and a small gap, the dot features a single orbital ground state and orbitally degenerate doublet of excited states, similar to the single-particle level spectrum of a two-dimensional harmonic oscillator. Increasing the dot and gap size in Fig.~\ref{fig_theo_exp}(c) enhances the effect of trigonal warping, leading to triplet degeneracies corresponding to the three minivalleys around each of bilayer graphene's valleys [see Fig.\,S2(e) in the Supplemental Material \cite{SM_all}] \cite{Knothe2020}.Note that, the parameters in Fig.~\ref{fig_theo_exp}(b) and (c) are chosen, such that we acount for the following effects: (1) the electric susceptibility of the bilayer's two monolayers, together wish the electron density redistribution between the layers, reduce the value of the gap compared to naive estimates \cite{Slizovskiy2019}; and (2) an increasing number of electrons inside the dot affect the shape of the confinement potential, causing it to be flatter and more shallow than that for the empty dot. In the Supplemental Material \cite{SM_all}, we show further dot spectra for a broader range of parameters, demonstrating that the change of the dot levels' multiplicity with gap and dot size is robust and does not depend on the exact choice of parameters.

Comparing the measured single-particle level spectrum in Fig.~\ref{fig_theo_exp}(a), where groups of four experimental data points correspond to one four-fold degenerate calculated data point in Figs.~\ref{fig_theo_exp}(b) and (c) (note that the first two data points in Fig.~\ref{fig_theo_exp}(a) overlap) with the theoretical model, the observed formation of the first two bunches of four and eight levels [see also Fig.~\ref{fig_Bperp}] agrees with the scenario in Fig.~\ref{fig_theo_exp}(b) of a small dot with a small gap. This scenario, however, does not predict the third bunch of twelve resonances observed in Fig.~\ref{fig_Bperp}; instead it foresees another two bunches of four and eight resonances, which is not observed. However, since both the displacement field and the quantum dot size increases with increasing electron number in the dot \cite{SM_all}, the large-dot-large-gap scenario depicted in Fig.~\ref{fig_theo_exp}(c) explains the bunching of twelve levels. This suggests a gradual transition from the scenario shown in Figs.~\ref{fig_theo_exp}(b) to that in Figs.~\ref{fig_theo_exp}(c) \cite{Knothe2020} while the quantum dot is being filled with more than five electrons. Such a transition is plausible for the following experimental reasons: First, charging energies extracted from finite-bias Coulomb-diamond measurements decrease consistently with increasing electron number \cite{SM_all}, indicating the increasing electronic size of the quantum dot. Second, given the negative back gate voltage in our experiment, an increasingly positive voltage $V_\mathrm{QD}$ on the finger gate increases the displacement field and thereby the induced band gap inside the quantum dot.

\begin{figure}
	\includegraphics[width=0.48 \textwidth]{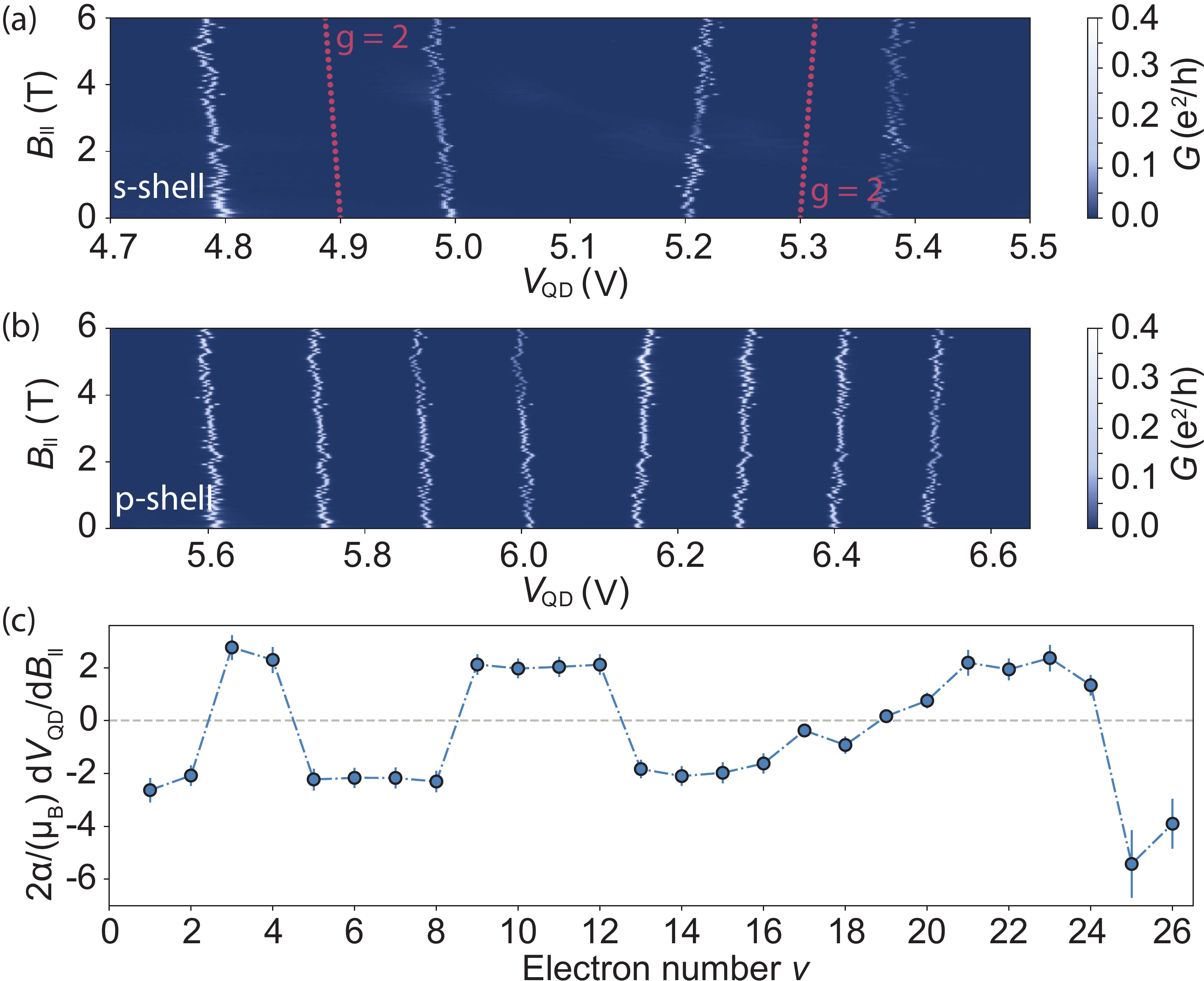}
	\caption{(a),(b) Conductance through the quantum dot with respect to parallel magnetic field $B_{\parallel}$ and $V_{\mathrm{QD}}$ for the first four and the fifth to twelfth electron, respectively. The red lines denote the slope of a Zeeman splitting with a g-factor of two for free electrons. (c) Fitted slopes of the shift of the resonances with magnetic field $\alpha / (2\mu_{B})\,\text{d} B_{\parallel} / \text{d}V_{\text{QD}} $. The energy levels $\nu$ are filled following Hund's second rule.}
	\label{fig2}
\end{figure}

Next we extract the spin-filling sequence of the quantum dot and the Zeeman splitting of the levels by performing measurements in magnetic field applied parallel to the graphene plane.
Figure~\ref{fig2}(a) shows the first bunch of four Coulomb peaks with parallel magnetic field. The dashed red lines provide a guide to the eye for a spin splitting with $g$-factor of $2$. In agreement with our earlier work \cite{Kurzmann2019}, we find in Fig.~\ref{fig2}(a) that the second electron is filled with its spin parallel to the first, indicating a two-electron spin-triplet and valley-singlet ground state. The third and fourth electrons are then filled with spin opposite to the first two, resulting in a filled shell with zero spin for four electrons in the dot. The observed spin-filling sequence can therefore be characterized by the total-spin quantum number sequence $s_z=1/2, 1, 1/2, 0$, which follows Hund's second rule.

Similarly, Fig.~\ref{fig2}(b) displays the magnetic field dependence of the fifth to twelfth conductance resonance corresponding to the second bunch of levels, i.e., the second shell. The first four electrons loaded into the quantum dot have the same spin, whereas the fifth to the eighth electrons are filled with the opposite spin. The total-spin quantum number sequence in this shell is therefore $s_z=1/2, 1, 3/2, 2, 3/2, 1, 1/2, 0$. The appearance of the remarkably high total spin $s_z=2$ for eight electrons in the dot shows that exchange interaction effects are stronger in this quantum dot than any level splittings induced by, for example, deviations from perfect circular symmetry or band non-parabolicities.

Figure~\ref{fig2}(c) shows the fitted slopes of shifts of the resonances in magnetic field
$2\alpha/\mu_\mathrm{B}\times \text{d}V_\mathrm{QD}/\text{d}B_{\parallel}$
which supports this interpretation. Furthermore, we extract g-factors from splittings of neighboring peaks \cite{SM_all, Lindemann2002}. For the first twelve electrons, the $g$-factor $|g|=2.3 \pm 0.3$ fits the expectation for electrons in graphene for small gap sizes and small dots. For larger electron numbers the situation is more complex. In Fig.~\ref{fig2}(c) we see that slopes tend to be smaller than the expected values of $\pm 2$, and there is a gradual change of slope from negative to positive values rather than an abrupt jump after the first six filled spins, which would correspond to a half-filled shell. While the exact origin of this behavior remains an open question, we speculate that exchange and correlation effects may play important roles in its explanation.

In summary, we performed measurements on a nearly circular dot, enabling us to observe the transition from filling $1\times 4$- and $2\times 4$-fold degenerate Fock-Darwin-like shells, to filling a $3\times 4$-fold degenerate shell governed by the three-fold minivalley symmetry. Observation of this transition was realized by increasing electron occupation of the dot, which is naturally accompanied with an increasing dot size and an increasing band gap.
We confirmed that the dot has a nearly circular shape by supporting electrostatic simulations of the potential landscape. Calculations of the single-particle level spectrum of a dot with circular symmetry are in qualitative agreement with our experimental results. Understanding the single-particle spectrum and its tunability is an important step towards identifying suitable states for qubit operation in bilayer graphene quantum dots.

\section{Acknowledgements}

We thank P. Märki and T. Bähler as well as the FIRST staff for their technical support. We thank B. Kratochwil for his support with data analysis.
We acknowledge funding from the Core3 European Graphene Flagship Project, the Swiss National Science Foundation via NCCR Quantum Science and Technology, the EU Spin-Nano RTN network, the European Quantum Technology Project 2D-SIPC, the ERC Synergy Grant Hetero2D, EPSRC grants EP/S030719/1 and EP/N010345/1 and the European Union’s Horizon 2020 research and innovation programme under the Marie Skłodowska-Curie Grant Agreement No. 766025.
Growth of hexagonal boron nitride crystals was supported by the Elemental Strategy Initiative conducted by the MEXT, Japan and JSPS KAKENHI Grant No. JP15K21722.

%

\newpage\null\newpage

\setcounter{section}{0} 

\renewcommand\thesection{S~\arabic{section}} 
\setcounter{figure}{0} 
\renewcommand\thefigure{S\arabic{figure}} 

\section{Finite bias measurements}

\begin{figure}
	\includegraphics[width=0.5 \textwidth]{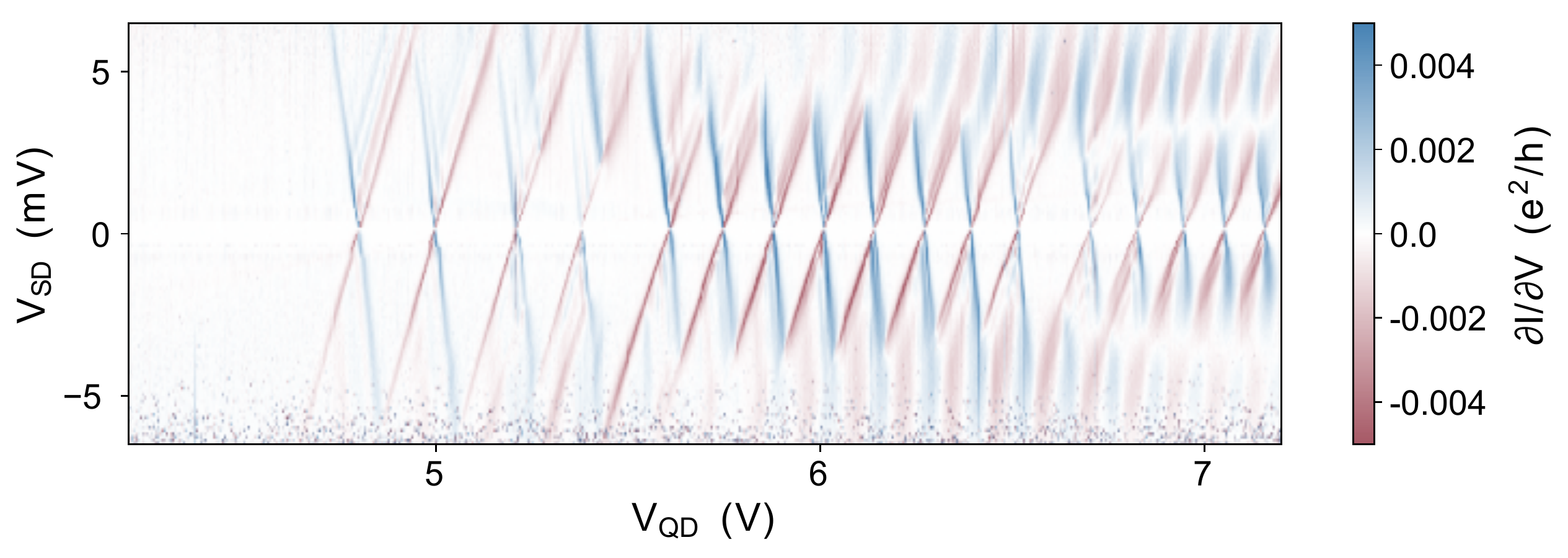}
	\caption{Finite bias spectroscopy $\partial I / \partial V_{SD}$ data. The size of the Coulomb diamonds gradually decreases with electron number, indicating an increase in dot size. The forth and twelfth diamonds are bigger than their neighbouring ones which shows the filling of one shell.}
	\label{diamond}
\end{figure}

To completely characterise the sample, we perform finite bias measurements for the first sixteen charge carriers in Fig. \ref{diamond}. We extract the lever arm for each charge carrier from the slopes of the resonances. The decreasing size of the diamonds indicates an increase in dot size for higher numbers of charge carriers. Furthermore, additional to the measurements presented in the main text, the coulomb diamonds also show the increase of the addition energy for the forth and twelfth charge carrier as the diamonds are larger than their neighbouring ones.

\section{Calculations on dot parameter}

\begin{figure}
	\includegraphics[width=0.5 \textwidth]{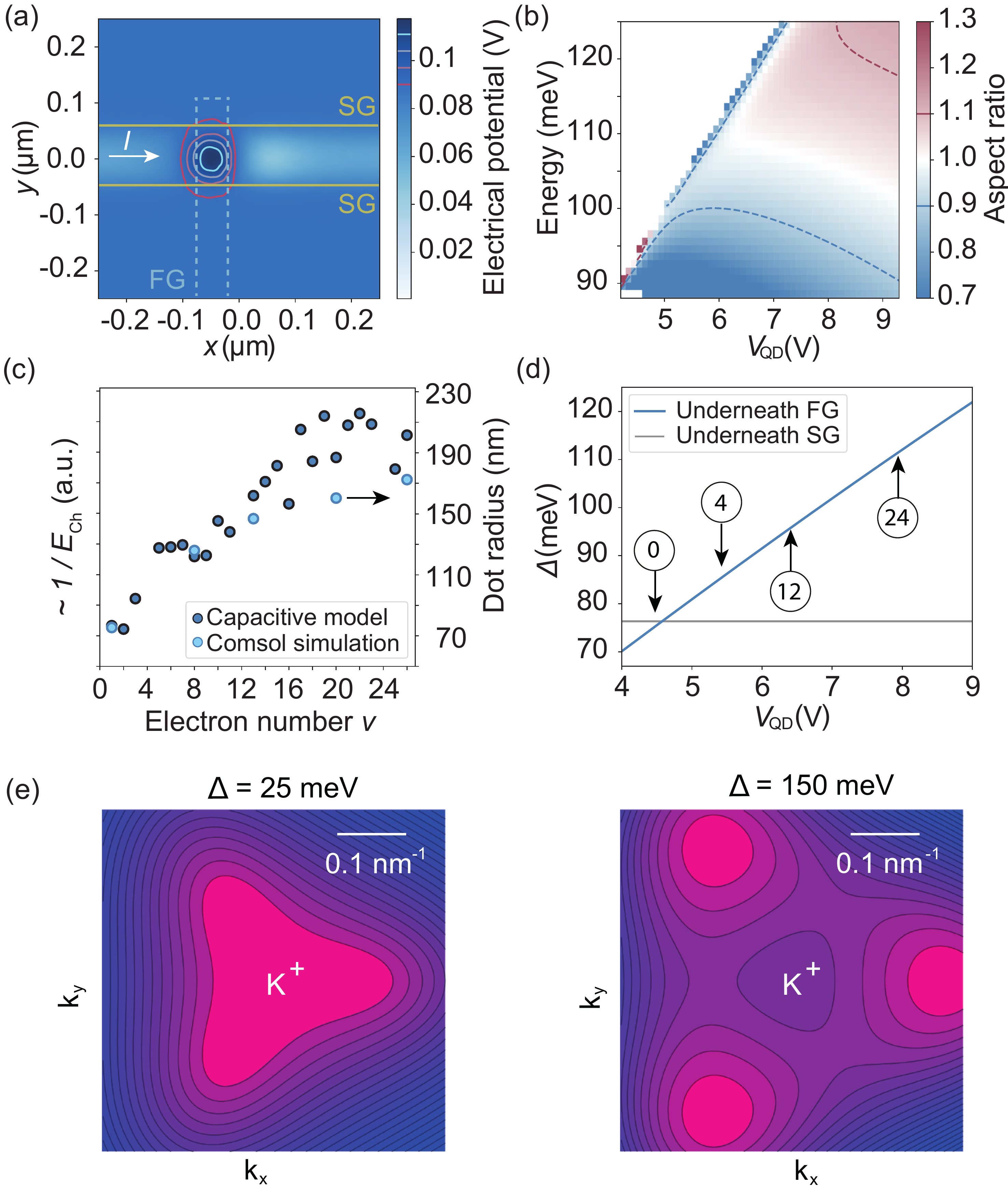}
	\caption{(a) Potential landscape induced to the bilayer graphene to form a quantum dot, where  $V_{\mathrm{QD}} = \SI{6.3}{V}$, calculated with \textsc{Comsol Multiphysics}. (b) Aspect ratio of the $x$- and $y$-axis of the dot, deduced from simulations as shown in (a) for different finger gate voltages $V_{\mathrm{QD}}$. The red and blue lines are a guide to the eye for an upper and lower limit of $\SI{10}{\percent}$ deviation of the perfect circle with an aspect ratio of one (white). (c) Dot radius for increasing number of electrons inside the dot deduced from the addition energy shown in Fig.~\ref{fig1}(d) (dark blue) and simulations (light blue). The dot radius is proportional to the charging energy. (d) Band gap opening in the bilayer graphene induced by the split gates and the finger gate with respect to the finger gate voltage $V_{\mathrm{QD}}$. The corresponding number of electrons in the dot is marked. (e) Low-energy dispersion (lowest conduction band) of gapped bilayer graphene around the $K^{+}$ valley for different sizes of the gap: larger gaps promote the formation of three clearly defined minivalleys around each valley. The color scale runs from magenta (energy at the band edge, lowest energy) to blue (high energy).}
	\label{fig_dot_param}
\end{figure}

We estimate the size and shape of the quantum dot by solving Poisson's equation in three dimensions with an electron density in the graphene plane determined self-consistently within the Thomas-Fermi approximation with \textsc{Comsol Multiphysics}. The gap size is estimated by using a tight binding model without self-screening of the graphene \cite{McCann2007}. Fig.~\ref{fig_dot_param}(a) shows the simulated potential landscape for $V_{\mathrm{QD}} = \SI{6.3}{V}$, which is the gate voltage for which the dot in the experiment is filled with ten electrons. The yellow lines indicate the split gates that form a conducting channel (light blue). The light blue dashed lines are a sketch of the finger gate that is used to form the dot. A higher electrical potential is observed below the finger gate, forming the potential for the quantum dot. The red, orange and blue lines each show a constant value for the electrical potential and are an indication for the shape of the quantum dot. 
Within a certain energy range, the equi-potentials in the well indeed have a nearly circular shape (e.g. light blue circle). We calculate the aspect ratio of the $x$- and $y$-axis of the dot for different voltages applied to the finger gate $V_{\text{QD}}$ and different energies [Fig.~\ref{fig_dot_param}(b)], where the energy cuts correspond to different contour lines of the potential landscape similar to Fig.~\ref{fig_dot_param}(a) to get a better estimation of the mentioned energy range, where the dot is circular. When the aspect ratio is between 1.1 and 0.9 we assume the dot to be round. The red and blue lines are a guide to the eye for this upper and lower limit of $\SI{10}{\percent}$ deviation of the perfect circle with an aspect ratio of one. It becomes clear that the dot can be considered round for a large parameter range in the cut-off energy (i.e. contour lines) as well as for $V_{\mathrm{QD}}$.

A rough estimate for the dot radius can be extracted from the addition energy, where the dot is assumed to be a circular disk-like capacitor of radius $r$, surrounded by a mixture of insulating hBN and amorphous Al$_2$O$_3$ with $\epsilon_r = 8.2$. Such a system has a self-capacitance of $C=8\epsilon_{\text{r}}\epsilon_0r$ \cite{Thomasbook}. Furthermore, the contribution of the confinement energy $\Delta = \hbar/4m^*r^2$ to the addition energy is included \cite{Eich2020}. The result is shown in Fig.~\ref{fig_dot_param}(c), starting at the first to the twelfth electron the dot radius becomes twice as large. For comparison, we extract the rough size of the dot from the simulated potential landscape (light blue data points in Fig.~\ref{fig_dot_param}(c)), which show the same trend.
Similarly to the split gates voltages $V_\mathrm{SG}$, $V_{\mathrm{QD}}$ also induces an increasing electrostatic gap in the bilayer graphene by increasing the displacement field. Assuming a parallel-plate capacitor model with the dielectrics hBN and Al$_2$O$_3$, but neglecting any self screening effect within the tight binding model, we calculate the size of the bandgap $\Delta$ depending on the finger gate voltage $V_{\mathrm{QD}}$ in Fig.~\ref{fig_dot_param}(d). The bandgap increased from about $\SI{75}{meV}$ when the quantum dot is charged with one electron, to $\SI{120}{meV}$, when the quantum dot is charged with $26$ electrons. These calculations give an upper bound estimate for the band gap induced in the bilayer graphene. A lower bound estimate of the gap induced at the first charge carrier of $\SI{30}{meV}$ can be deducted from the separation between the pinch-off to zero conductance and the first Coulomb resonance in a conductance trace similar to the one shown in Fig.\,1(c).

\section{Theoretical model of the quantum dot}
\label{subsection:theoryModel}
\begin{figure*}[t]
	\includegraphics[width=1 \textwidth]{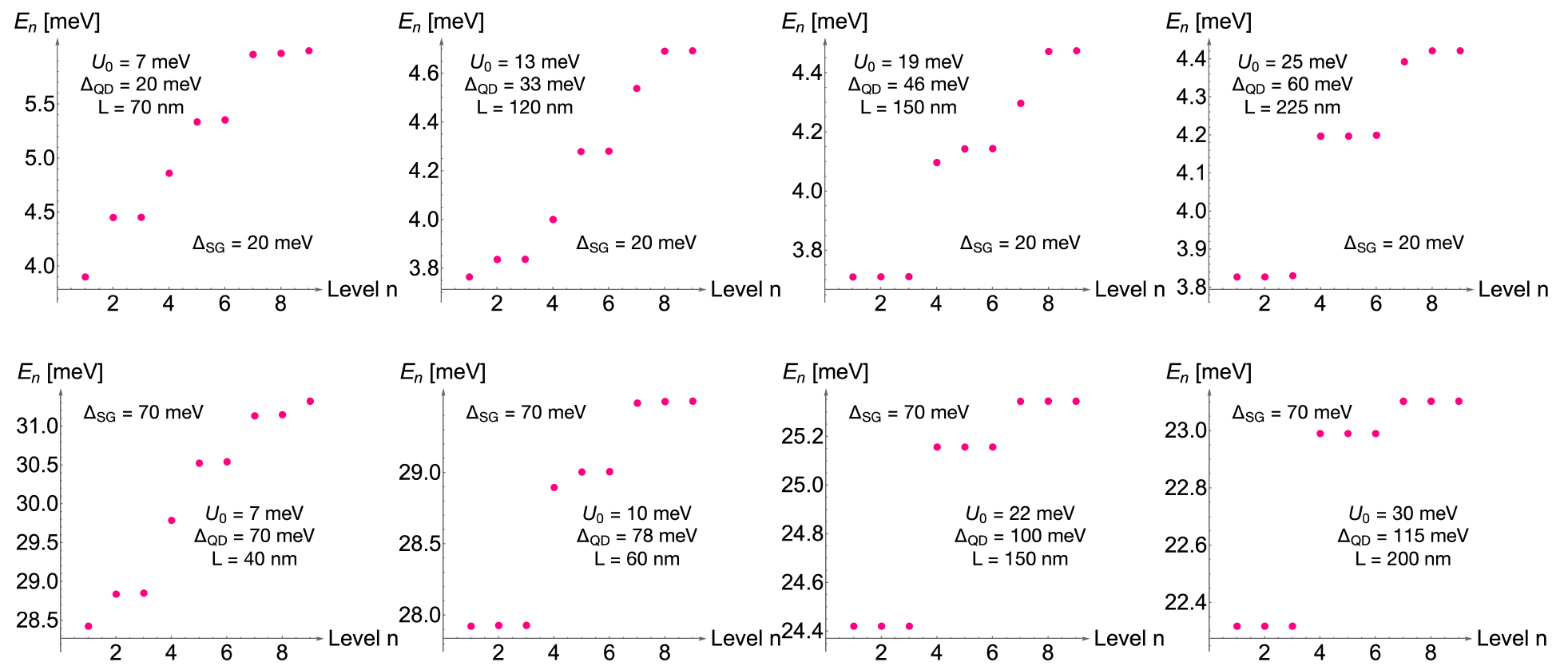}
	\caption{Bilayer gaphene quantum dot spectra. In each row, the dot size and the gap inside the dot increase from left to right (controlled by the increasing finger gate voltage in the experiment), while the gap underneath the split gates is held constant. With increasing dot and gap size the multiplicity of the levels changes and orbital triplet degeneracies emerge as a result of bilayer graphene's three minivalleys around each K-point. Top and bottom row are for different initial values of the dot parameters and the split gate gap, demonstrating that the change in the dot spectras' multiplicity is a robust feature over a large parameter range.}
	\label{fig:Spectra}
\end{figure*}

We describe the circularly rotational dot in bilayer graphene by a smooth confinement potential, $U$, and spatially varying gap, $\Delta$, with functional dependences,
\begin{align}
    \nonumber U(\mathbf{r})&=\frac{U_0}{\cosh{\frac{r}{L}}}, \\
    \Delta(\mathbf{r})&=\Delta_{SG}+\frac{\Delta_{mod}}{\cosh{\frac{r}{L}}},
\end{align}
with $r=|\mathbf{r}|$ and $\mathbf{r}=(x,y)$, and dot parameters, $U_0$, $L$, a constant part of the gap, $\Delta_{SG}$, and a modulated gap component, scaled by $\Delta_{mod}$. These potentials enter in the four-band Hamiltonian of bilayer graphene \cite{McCann2007, McCann2013},
\begin{equation}
    H_{\pm}=
\setlength{\arraycolsep}{+5pt} \begin{pmatrix} 
 U \mp\frac{1}{2}\Delta  & \pm v_3\pi & 0 &\pm v \pi^{\dagger}\\
\pm v_3 \pi^{\dagger}&  U \pm\frac{1}{2}\Delta  & \pm v\pi &0\\
 0 & \pm v\pi^{\dagger} &   U \pm\frac{1}{2}\Delta  &   \gamma_1\\
\pm v\pi & 0 &   \gamma_1 &  U \mp\frac{1}{2}\Delta 
\end{pmatrix},
\label{eqn:H}
\end{equation}
for the two valleys $K^{\pm}$, with momenta $\pi=p_x+ip_y,\,  \pi^{\dagger}=p_x-ip_y$,  velocities $v=1.02*10^6 \text{ m/s}$ and  $v_3\approx0.12 v$, and energy $\gamma_1\approx0.38\text{ eV}$. The Hamiltonian  is written in the basis $\psi_{K^+}=(\psi_{A},\psi_{B^{\prime}},\psi_{A^{\prime}},\psi_{B})$ in valley $K^+$, and $\psi_{K^-}=(\psi_{B^{\prime}},\psi_{A},\psi_{B},\psi_{A^{\prime}})$ in valley $K^-$,  with electron's amplitudes on the bilayer graphene sublattices  $A$ and $B$ ($A^{\prime}$ and $B^{\prime}$) in the top (bottom) layer. 
In the absence of confinement, Eq.~\eqref{eqn:H} describes the low energy trigonally warped bands of bilayer graphene \cite{Varlet2014, Varlet2015, Knothe2018, Knothe2020},  featuring three minivalleys around each $K$ point.
The larger the bias, the better the minivalleys are defined, becoming deeper and more separated in momentum space. \\
In the presence of confinement, we diagonalize the Hamiltonian,  $H_{\pm}$, in Eq.\eqref{eqn:H} numerically in a basis of localised states  \cite{Knothe2018, Knothe2020, tong2020}. We choose the eigenstates of the two-dimensional  harmonic oscillator  (products of wave functions $\psi_n(x)=N_n e^{-\frac{1}{2}(\alpha x)^2}\mathcal{H}_n(\alpha x)$, where $N_n= \sqrt{\frac{\alpha}{\sqrt{\pi}2^n n!}}$ is the normalization constant and $\alpha$ is a scaling factor of unit length$^{-1}$ adapted to the bottom of $U$). The basis states are  given by 
\begin{equation}\psi_{\eta\mu ,1}=
\nonumber\begin{pmatrix}
\psi_{\eta}(x)\psi_{\mu}(y)\\
0\\
0\\
0
\end{pmatrix},\;
\psi_{\eta\mu,2}=
\begin{pmatrix}
0\\
\psi_{\eta}(x)\psi_{\mu}(y)\\
0\\
0
\end{pmatrix},
\end{equation}
\begin{equation}
\psi_{\eta\mu,3}=
\begin{pmatrix}
0\\
0\\
\psi_{\eta}(x)\psi_{\mu}(y)\\
0
\end{pmatrix},
\psi_{ \eta\mu,4}=
\begin{pmatrix}
0\\
0\\
0\\
\psi_{\eta}(x)\psi_{\mu}(y).
\end{pmatrix}.
 \label{eqn:BasisHarm}
\end{equation}
For every set of system parameters we construct the matrix corresponding to Hamiltonian $H_{\pm}$ in the basis of Eqn.~\eqref{eqn:BasisHarm}  and obtain the dot's energy spectrum by diagonalization. We reach convergence when the energy levels do not change further upon including a higher number of basis states.

\subsection*{Dot spectra over a large parameter range}
\label{subsection:theorySpectra}
In the experiment, the dot's width and depth, $L$ and $U_0$, are varied together with the gap size inside the dot, $\Delta_{QD}=\Delta_{SG}+\Delta_{mod}$, all of which increase with increasing finger gate voltage. The constant gap underneath the split gates, $\Delta_{SG}$, is held fixed. In Fig.~\ref{fig:Spectra}, we illustrate the evolution of the dot spectra for different system parameters and increasing  $L$  $U_0$, and $\Delta_{QD}$. For small dots and small gaps, the dot's ground state is singly orbitally degenerate, and the level structure resembles that of parabolic confinement. With increasing dot and gap size, the orbital multiplicity of the levels changes, and triplet degeneracies emerge, corresponding to degenerate states in the three minivalleys connected by $\frac{\pi}{3}$ rotation. For sufficiently large gaps and dots, all levels come three-fold degenerate. Larger bias, larger dots, and flatter confinement potentials facilitate the formation of triplet states.

\section{Evaluation of $g$-factor}
\label{subsection:Lindemann}

\begin{figure*}
	\includegraphics[width=1 \textwidth]{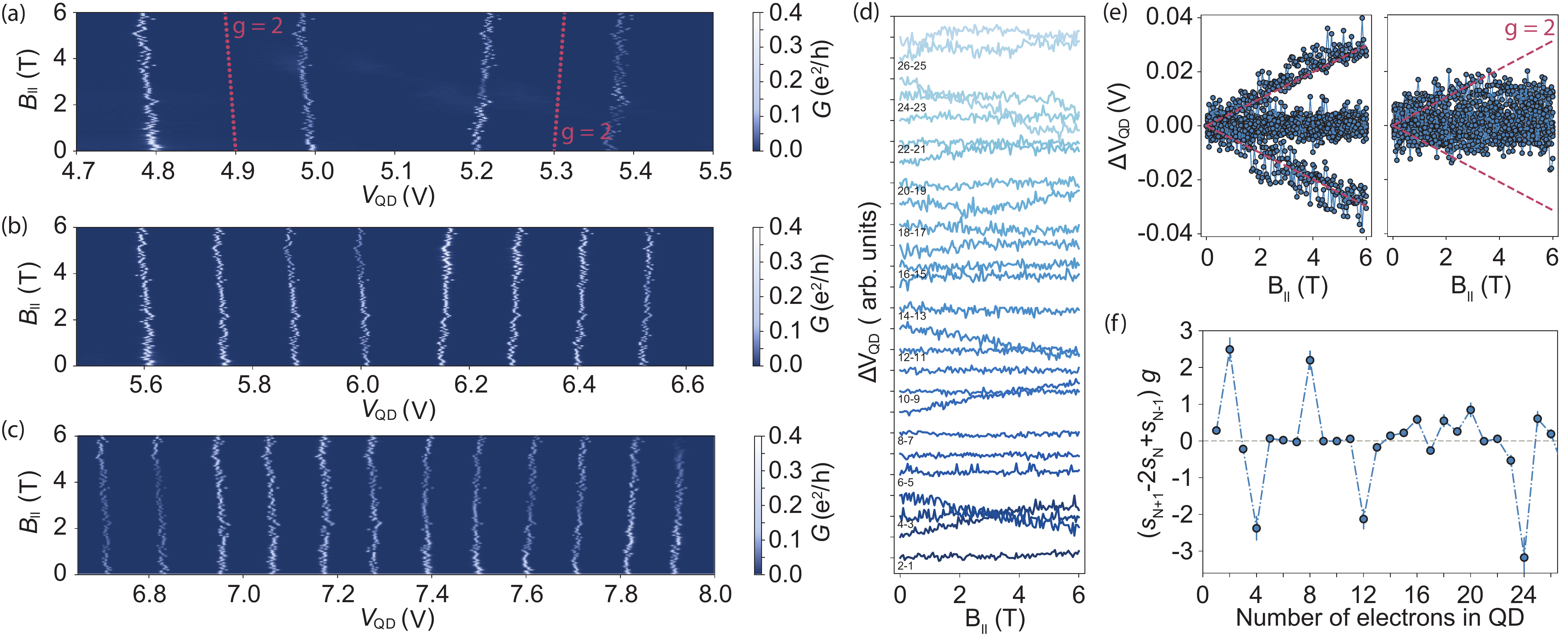}
	\caption{(a)-(c) Conductance through the quantum dot with respect to parallel magnetic field $B_{\parallel}$ and $V_{\mathrm{QD}}$ for the first four, fifth to twelfth and thirteenth to twenty-fourth electron, respectively. The red lines denote the slope of a Zeeman splitting with a g-factor of two for free electrons. (d) Evolution of peak-spacing with magnetic field. The peak spacing is extracted from the conductance traces in (a)-(c) with a vertically offset for clarity. (e) Peak spacings offset to align spacings at zero field. The red dashed lines show the slope corresponding to a $g$ factor $g=2$. Left: Peak spacings one to twelve. Right: Peak spacings thirteen to twenty-four. (f) Change of the spin occupancy inside the quantum dot times the $g$-factor versus the number of electrons inside the dot.}
	\label{Fig_Lindemann}
\end{figure*}

Since the spin-orbit coupling in bilayer graphene structures is negligible the total energy of the N-electron quantum dot state in a parallel magnetic field $B$ is the sum of an orbital contribution and a spin contribution yielding \cite{Lindemann2002}
\begin{eqnarray*}\lefteqn{e \alpha_{N+1}\Delta V_{\text{QD}}^{N+1}(B_{\parallel}) =} \\
& & (s_{N+1}-2s_N+s_{N-1})g\mu_BB_{\parallel}+\text{const}.
\end{eqnarray*}

Here $s_N$ is the component along the direction of $B$ of the total spin of the quantum dot with $N$ electrons, $\alpha_N$ is the respective lever arm and $g$ is the magnitude of the Zeeman splitting which is expected to be $g=2$ in graphene. Plotting $\Delta V_{\text{QD}}$ versus $B_{\parallel}$ shows branches with slopes proportional to $0, \,\pm g,\,\pm 2g,...$, where slopes proportional to $\pm 2g,...$ require a spin flip and are not observed in our measurements. Fig.~\ref{Fig_Lindemann}(a)-(c) show the behaviour of the conductance resonances in a parallel magnetic field for the first, second and third bunch of resonances respectively. The consecutive splitting of the resonances is extracted and presented with an offset for better visibility in Fig.~\ref{Fig_Lindemann}(d). The corresponding number of the subtracted resonances are labeled. Subtracting the offset at zero field for each peak spacing yields Fig.~\ref{Fig_Lindemann}(e), where the first twelve spacings are presented in the left panel and peak spacing thirteen to twenty-four in the right panel. The red lines provide a guide to the eye of a $g$-factor of $g=2$. Extracting the slope of every peak spacing with magnetic field, multiplying it with the respective lever arm $\alpha$ from finite bias measurements and divided by $\mu_B$ yields the change of spin occupancy inside the quantum dot multiplied by $g$. We extract a $g$-factor for the Zeeman splitting of $g=2.3 \pm 0.3$ for the first twelve electrons in the quantum dot. For higher numbers of electrons in the quantum dot, the splitting with magnetic field deviates from classic Zeeman splitting and cannot be explained with this simple model anymore.

\end{document}